\documentclass[square]{ws-procs975x65}

\newcommand{\be}{\begin{equation}}
\newcommand{\ee}{\end{equation}}
\begin{document}

\title{THE ENERGY RELEASE--STELLAR ANGULAR MOMENTUM 
INDEPENDENCE IN ROTATING COMPACT STARS UNDERGOING FIRST-ORDER PHASE 
TRANSITIONS}

\author{M. BEJGER$^{1,2}$, J. L. ZDUNIK$^2$, P. HAENSEL$^2$ and E. GOURGOULHON$^1$}
\address{$^1$ LUTh, Observatoire de Paris, CNRS, Universit\'e Paris Diderot,
5 Place Jules Janssen, 92190 Meudon, France\\
$^2$ N. Copernicus Astronomical Center, Polish 
Academy of Sciences, Bartycka 18, PL-00-716 Warszawa, Poland\\
bejger@camk.edu.pl, jlz@camk.edu.pl, haensel@camk.edu.pl, Eric.Gourgoulhon@obspm.fr}

\begin{abstract}
We present the general relativistic calculation of the energy 
release associated with a first order phase transition (PT) at the 
center of a rotating neutron star (NS). The energy release, $E_{\rm rel}$, is 
equal to the difference in mass-energies between the initial 
(normal) phase configuration and the final configuration 
containing a superdense matter core, assuming constant total 
baryon number and the angular momentum. The calculations are
performed with the use of precise pseudo-spectral 2-D numerical code; 
the polytropic equations of state (EOS) as well as realistic EOSs 
(Skyrme interactions, Mean Field Theory kaon condensate) are used. 
The results are obtained for a broad range of metastability
of initial configuration and size of the new superdense phase core 
in the final configuration. For a fixed ``overpressure'', $\delta\overline{P}$,
defined as the relative excess of central pressure of a collapsing
metastable star over the pressure of the equilibrium
first-order PT, the energy release up to numerical accuracy 
{\it does not depend} on the stellar angular momentum and coincides 
with that for nonrotating stars with the same $\delta\overline{P}$.
When the equatorial radius of the superdense phase core is much smaller 
than the equatorial radius of the star, analytical expressions for 
the $E_{\rm rel}$ can be obtained: $E_{\rm rel}$ is proportional to 
$(\delta\overline{P})^{2.5}$ for small $\delta\overline{P}$. 
At higher $\delta\overline{P}$, the results of 1-D calculations 
of $E_{\rm rel}(\delta\overline{P})$ for non-rotating stars reproduce with very
high precision exact 2-D results for fast-rotating stars.
The energy release-angular momentum independence for a given overpressure 
holds also for the so-called ``strong'' PTs 
(that destabilise the star against the axi-symmetric perturbations),
as well as for PTs with ``jumping'' over the energy barrier.
\end{abstract}

\bodymatter

\section{Introduction}\label{introduction}
Many theories of dense matter predict that at some density larger than 
the nuclear saturation density, a phase transition (PT) to 
some ``exotic'' state (i.e. boson condensate or quark deconfinement) occurs; 
for review see e.g., \cite{NS1,Glend.book,weber99}). A first-order PTs 
are particularly interesting from the astrophysical and observational point of view, 
because are associated with a meta-stable state of dense matter. One can thus expect, 
in the case of PTs occurring in the interior of NSs, the release of non-negligible 
amount of energy. Here we will focus on the basic features of the energy 
release-angular momentum independence; for complete description we refer the reader 
to \cite{erot,erots}. The text is arranged as follows: in Sect.~\ref{calculations}
we briefly describe the results of calculations. Sect.~\ref{discussion} contains 
conclusions and open questions.

\begin{figure}[t]
\begin{center}
\psfig{file=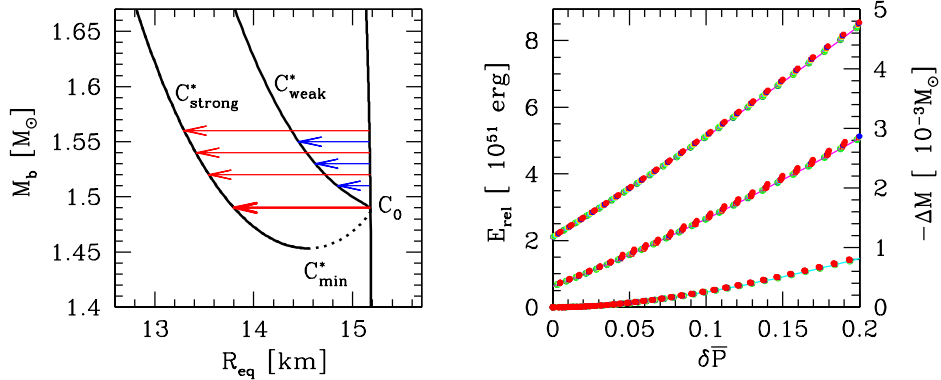,width=5.in}
\end{center}
\caption{
Left panel: Strong and weak PTs on mass-radius diagram (dotted line 
marks unstable configurations). 
Right panel: the energy release $E_{\rm rel}$ vs the over-pressure parameter
$\delta\bar{P}$ for strong (two upper curves) and weak (lowest curve)
PTs. Points are colored differently for different total angular momenta 
$J=(0,0.1,...,1.3)\times GM^2_\odot/c$ and follow the curve for the $J=0$
 (non-rotating) configurations.}
\label{reffig1}
\end{figure}

\section{Calculation of the energy release}
The hydrostatic, axi-symmetric and rigidly rotating configurations of compact stars 
with and without PTs were obtained  using the numerical GR library {\sc LORENE} 
({\tt http://www.lorene.obspm.fr}). We assume that the baryon mass $M_{\rm b}$ and 
the total angular momentum $J$ are constant during the PT from the metastable 
configuration $\cal{C}$ to stable configuration $\cal{C}^*$. From the microscopic 
point of view the first-order PT is characterised by the over-pressure parameter
$\delta\bar{P}=(P_c-P_0)/P_0$, where $P_c$ is the central pressure of the configuration 
${\cal C}$ and $P_0$ is the equilibrium pressure, at which the PT occurs. 
We distinguish between the so-called {\it weak} and {\it strong} PTs, 
characterised by the density-jump parameter $\lambda=\rho^*/\rho$, where $\rho$ and 
$\rho^*$ are the densities of normal and condensed phase at $P_0$. 
PTs with $\lambda > {\frac{3}{2}}(1+P_0/\rho c^2)$ (strong PTs) destabilise the stars 
with arbitrarily small cores against the axi-symmetric perturbations
\cite{Seidov1971,Kaempfer1981,ZdunikHS1987}. The energy release (energy difference) 
is defined as 
\be
E_{\rm rel}=c^2[M({\cal C})-M({\cal C}^*)]_{M_{\rm b},J}
\ee
and it is quite remarkable that (up to numerical accuracy) it does not depend 
on the rotation state of the configuration i.e. on the 
total angular momentum $J$ of the star and agrees well 
with the $E_{\rm rel}$ for non-rotating stars. In Fig.~\ref{reffig1} we present 
the results for the polytropes (parametric EOSs in the form of $P=Kn^\Gamma$, where 
$n$ is the number baryon density, $K$ is is the pressure coefficient and $\Gamma$ 
is called the adiabatic index; detailed description of parameters used can be found 
in \cite{erots}).  

\begin{figure}[t]
\begin{center}
\psfig{file=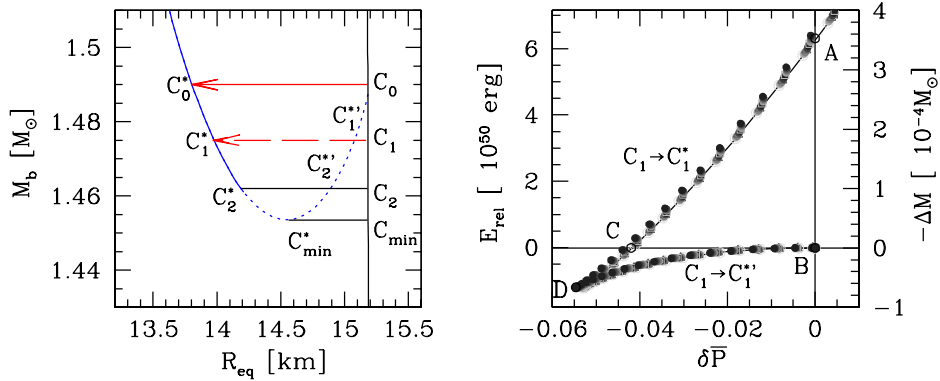,width=5.in}
\end{center}
\caption{Left panel: The strong PT with the ``jumping'' over the energy barrier 
on the mass-radius diagram. Right: Energy release $E_{\rm rel}$ vs the over-pressure 
parameter $\delta\bar{P}$ for the strong PT. Points with different colors correspond
to PTs with different angular momentum
$J=(0, 0.1,\dots, 0.9)\times GM_\odot^2/c^2$ also for negative $\delta\bar{P}$. 
Solid line - result for $J=0$ (non-rotating stars). 
Point A corresponds to ${\cal C}_0\longrightarrow {\cal C}_0^\star$, 
point B to ${\cal C}_0\longrightarrow {\cal C}^{\star\prime}_0={\cal C}_0$, 
point C  to ${\cal C}_2\longrightarrow {\cal
C}_2^\star$ and point D to ${\cal C}_{\rm min}\longrightarrow {\cal C}_{\rm
min}^\star$.}
\label{reffig2}
\end{figure}

In the case of realistic EOSs the results are qualitatively the same. We have checked 
the phenomenon for different realistic EOSs i.e. Skyrme interactions or the kaon 
condensate, presented as an example here. 
The crust obeys the EOS of Douchin \& Haensel\cite{DouchinH2001}. 
The matter below the PT is described using the relativistic mean-field theory, 
\cite{ZimanyiM1990}. The dense phase is the kaon condensate, with coupling of kaons to nucleons proposed by Glendenning \& Schaffner-Bielich\cite{GlendenningS1999}, the optical potential $U^{\rm lin}_{\rm K}$ being equal to $-115$ MeV. 
The stability of configurations with $P_c$ 
below that for the equilibrium PT is particularly interesting. 
If the star is excited initially, e.g. is pulsating, 
then the formation of a large dense phase core is
possible, but it requires climbing (``jumping'') over the 
energy barrier associated with formation of a small core. 
The results are shown in Fig.~\ref{reffig2}.
\label{calculations}

\section{Conclusions and open questions}
\label{discussion}
Our numerical calculations show the $E_{\rm rel}(J)$ independence 
during the weak as well as strong first-order PTs, for large rotation rates and 
large oblatnesses of the stars: it is therefore not an property of slow rotating 
stars only. The independence holds also for 
PTs with negative over-pressure $\delta\bar{P}$ i.e. when the configuration 
``jumps'' over the energy barrier to reach another stable configuration. 
For small positive $\delta\bar{P}$ analytical relations were found: 
$E_{\rm rel} \propto (\delta\overline{P})^{2.5}$. 
The energy release $E_{\rm rel}\sim 10^{51}-10^{52}$ erg is
an absolute upper bound on the energy which can released in such PT.
In astrophysical situation, the energy can be distributed between stellar pulsations,
gravitational radiation, heating of stellar interior, X-ray
emission from the neutron star surface, and even a gamma-ray burst. 

Currently there is no mathematical proof of the $J$-independence 
of $E_{\rm rel}$. 
It may be interesting to look at the problem from the ``thermodynamical'' 
point of view. We write the total energy of the star (gravitational mass $M$) as
\be \label{e:M_A_J_p}
        M = \frac{\mu}{u^t} A + 2 \, \Omega J
        + 2 \int_{\Sigma_t} P\,  N\sqrt{\gamma}\,  d^3x ,
\ee
where $\mu$ is the baryon chemical potential, $u^t$ the time component of 
the fluid 4-velocity, $A$ the number of baryons in the star, 
$\Omega$ the angular velocity, $P$ the fluid pressure, 
$N$ the lapse function and $\sqrt{\gamma}\, d^3x$ the covariant volume element 
in the constant $t$ hypersurface $\Sigma_t$.
Eq.~(\ref{e:M_A_J_p}) is valid for any axi-symmetric stationary and rigidly 
rotating fluid star which obeys a barotropic EOS, as established by 
Bardeen \& Wagoner in 1971 \cite{BardeW71}. 
Since $\mathcal{C}$ and $\mathcal{C}^*$ have the same baryon number $A$ 
and the same angular momentum $J$, we get the energy release 
\be \label{e:DE}
 E_{\rm rel} = \Delta E = A \, \Delta\! \left( \frac{\mu}{u^t} \right)
        + 2\,  J \, \Delta\Omega
        + 2 \left[ \int_{\mathcal{C}} P\,  N\sqrt{\gamma}\,  d^3x
        - \int_{\mathcal{C}^*} P\,  N\sqrt{\gamma}\,  d^3x \right] ,
\ee
with
\be
 \Delta\! \left( \frac{\mu}{u^t} \right) :=
        \left. \frac{\mu}{u^t} \right| _{\mathcal{C}} -
        \left. \frac{\mu}{u^t} \right| _{\mathcal{C}^*} ,~~~
 \Delta \Omega := \left. \Omega \right| _{\mathcal{C}}
        - \left. \Omega \right| _{\mathcal{C}^*} .
\ee
The terms on the right-hand-side of Eq.~\ref{e:DE} are of comparable magnitude 
in their contribution 
to the energy release, so the possible ``smallness'' of some 
of them in relation to the others is not responsible for the $E_{\rm rel}(J)$ 
independence. We will address the problem of mathematically proving 
this property in the near future. 
\section*{Acknowledgements}
This work was partially supported by
the Polish MNiI grant no. 1P03D.008.27, MNiSW grant no.
N203.006.32/0450 and by the LEA Astro-PF programme. 
MB was also supported by the Marie Curie Fellowship no. MEIF-CT-2005-023644.
% within the 6th European Community Framework Programme.
\bibliographystyle{ws-procs975x65}
\bibliography{bejger}
\end{document}